\documentclass[11pt,superscriptaddress,aip,preprint]{revtex4}
\usepackage[dvips]{graphicx}
\usepackage{amsfonts}
\usepackage{slashed}
\usepackage{amsmath}

\newcommand{\sech}{\mathrm{sech}}
\newcommand{\eff}{\mathrm{eff}}
\newcommand{\Tr}{\mathrm{Tr}}
\newcommand{\tr}{\mathrm{tr}}

\begin{document}

\title{Induced higher-derivative Lorentz-violating Chern-Simons term at finite temperature}

\author{J. Leite}
\affiliation{Instituto de F\'\i sica, Universidade Federal de Alagoas, 57072-270, Macei\'o, Alagoas, Brazil}
\email{julioleite,tmariz,wserafim@fis.ufal.br}

\author{T. Mariz}
\affiliation{Instituto de F\'\i sica, Universidade Federal de Alagoas, 57072-270, Macei\'o, Alagoas, Brazil}
\email{julioleite,tmariz,wserafim@fis.ufal.br}

\author{W. Serafim}
\affiliation{Instituto de F\'\i sica, Universidade Federal de Alagoas, 57072-270, Macei\'o, Alagoas, Brazil}
\email{julioleite,tmariz,wserafim@fis.ufal.br}

\date{\today}

\begin{abstract}
In this work, we analyze the generation of the higher-derivative Lorentz-violating Chern-Simons term at zero temperature and at finite temperature. We use the method of derivative expansion and the Matsubara formalism in order to consider the finite temperature effects. The results show that at zero temperature the induced higher-derivative Chern-Simons term is nonzero; in contrast, when the temperature reaches infinity, the coefficients of the induced term vanish. In addition, we also briefly study the question of large gauge invariance of this higher-derivative term as we as the conventional Chern-Simons term. We compute the exact induced action for both terms at finite temperature, however, in a particular gauge field background, and observe that they are, in fact, invariant under large gauge transformation.
\end{abstract}

\maketitle

\section{Introduction}
\label{intro}

Lorentz symmetry is one of the most tested symmetries of nature. Nevertheless, it is nowadays believed not to be an exact symmetry. In fact, it has been argued in~\cite{Kos1} that tiny spontaneous breaking of Lorentz symmetry might arise in the context of string theory. Thus, if we add Lorentz-violating terms to the standard model, this new theory, namely the standard-model extension (SME)~\cite{Kos2,Kos3}, can be viewed as the low-energy limit of a physically relevant fundamental theory in which spontaneous Lorentz violation occurs. Many experiments and observations have been conducted to detect physical effects due to the Lorentz violation and some constraints on the Lorentz-violating terms were obtained, as we can see in~\cite{Kos4}.

The most studied part of the SME presents renormalizable operators of mass dimension $d=3$ and $d=4$, consequently, the coefficients contracted with these operators have mass dimension $d=1$ or they are dimensionless, respectively. This part of the SME is known as the minimal SME. However, in the SME there are operators with mass dimension $d \geq 5$, which are nonrenormalizable, contracted with coefficients of mass dimension $d\leq -1$. Although these operators are likely to be specially relevant in searches involving very high energies, providing new corrections, there are few studies on them. 

Recently, some studies on higher-derivative Lorentz-violating terms, described by higher mass dimension operator ($d\geq 5$), have been made, such as~\cite{Gam,Alf,Sud,Mye,Bol,Jac,Bol1,Rey,Kos5,Rey2,Mar,Lei,Kos6,Bri}. In particular, the work~\cite{Kos5} on operators of arbitrary mass dimension in the QED extension have shown that the general form of the higher-derivative terms in the photon sector, is given by
\begin{eqnarray}\label{hdd}
S_{(d)}=\int{d^4x}\, \mathcal{K}^{\alpha_1 \alpha_2 \cdots \alpha_d}_{(d)} A_{\alpha_1}\partial_{\alpha_3}\cdots \partial_{\alpha_d} A_{\alpha_2},
\end{eqnarray}
where $d$ is the dimension of the tensor operator and $\mathcal{K}^{\alpha_1 \alpha_2 \cdots \alpha_d}_{(d)}$ has mass dimension $4-d$. 

On the one hand, for CPT-even terms, the first four indices of the coefficient $\mathcal{K}^{\alpha_1 \alpha_2 \cdots \alpha_d}_{(d)}$ have the symmetries of the Riemann tensor, and there is total symmetry in the remaining $d-4$ indices. On the other hand, for CPT-odd terms, the coefficient $\mathcal{K}^{\alpha_1 \alpha_2 \cdots \alpha_d}_{(d)}$ is antisymmetric in the three first indices and symmetric in the last $d-3$. 

In this work, we study the fermion sector of the quantum electrodynamics (QED) extended by a CPT-odd term, governed by the coefficient $b_\mu$, namely $b_\mu\bar{\psi}\gamma_5\gamma^\mu\psi$. From the fermion sector of this minimal QED extension, at zero temperature, we know that we can generate through radiative corrections the higher-derivative Chern-Simons term,
\begin{eqnarray}\label{hd5}
S_{(5)}=\int{d^4x}\, \mathcal{K}^{\mu \nu \rho \alpha \beta}_{(5)} A_{\mu}\partial_{\rho} \partial_{\alpha}\partial_{\beta} A_{\nu},
\end{eqnarray}
with $\mathcal{K}^{\mu \nu \rho \alpha \beta}_{(5)}\propto \epsilon^{\mu \nu \rho \sigma} b_\sigma g^{\alpha \beta}$, as can be seen by expanding the results in~\cite{Kos3,JacKos} or directly in~\cite{Bon,Mar2}. Our goal here is to calculate the generation of the above term (\ref{hd5}) using the method of derivative expansion \cite{Ait,Fra,Vai,Bab}, as a preparation for the finite temperature analysis, which has never been studied. As a complement, we also address the question of large gauge invariance of this higher-derivative term as well as the conventional Chern-Simons term, both at finite temperature. For issues related to the Chern-Simons term, ${\cal L}_{(3)} =\mathcal{K}^{\mu \nu \rho}_{(3)} A_{\mu}\partial_{\rho} A_{\nu}$, see \cite{Col,Chu,Per,Cha,Ada,And,Bel,Mar3}, and references therein.

It is interesting to mention that the term (\ref{hd5}) arises when the Myers-Pospelov bosonic term \cite{Mye} is radiatively induced from the fermion sector \cite{Mar2}. Therefore, effects of finite temperature on the higher-derivative Chern-Simons term are somehow related to the effects on the Myers-Pospelov term. Thus, studies of this nature deserve to be considered in order to better characterize the higher-derivative Lorentz-violating models. 

This paper has the following structure. In Sec.~\ref{zero}, we analyze the fermion sector of the Lorentz-violating QED with the coefficient $b_\mu$, performing radiative corrections and using derivative expansion to induce the higher-derivative Chern-Simons term at zero temperature. In the end of the section, we look for constraints on the coefficient of the induced term imposed by experimental and observational data. In Sec.~\ref{finite}, we keep looking at the same theory of the previous section, but now we are interested on the behavior of the term (\ref{hd5}) at finite temperature. The issue of large gauge invariance is discussed in Sec.~\ref{large}. Finally, we present a summary in Sec.~\ref{summary}.

\section{Radiative correction at zero temperature}
\label{zero}

As our aim is to study the generation of higher-derivative CPT-odd terms (\ref{hd5}), we consider the fermion sector of the Lorentz-violating QED, given by
\begin{eqnarray}\label{lg}
\mathcal{L}=\bar{\psi}(i\slashed{\partial}-m-\slashed{A}-\gamma_5\slashed{b})\psi,
\end{eqnarray} 
where the three first terms are the usual ones, and the last and unusual term, which is governed by the coefficient $b^\mu$, is that responsible for the Lorentz and CPT violation. 

In order to perform this task, we prefer to use the method of derivative expansion (for details see~\cite{Baz}), which is more suitable to calculate the sum-integrals of the finite-temperature approach.

Then, from (\ref{lg}), the effective action $S_{\eff}$ is defined as
\begin{eqnarray}
Z = \int D \bar{\psi} D \psi e^{i\int d^4 x \bar{\psi}(i\slashed{\partial}-m-\slashed{A}-\gamma_5\slashed{b})\psi} = e^{i S_{\eff}},
\end{eqnarray}
in which 
\begin{equation}\label{seff}
S_{\eff}=-i\Tr\ln(\slashed{p}-m- \slashed{A}-\gamma_5 \slashed{b}).
\end{equation} 
Now, we can easily rewrite the effective action as $S_{\eff}=S^{(0)}_{\eff}+S^{(1)}_{\eff}$, where $S_\eff^{(0)}=-i \Tr \ln{(\slashed{p}-m- \gamma_5 \slashed{b})}$ and 
\begin{eqnarray}\label{seff1}
S^{(1)}_{\eff}=i \Tr \sum_{n=1}^{\infty}\frac{1}{n}\left[\frac{1}{\slashed{p}-m-\gamma_5\slashed{b}} \slashed{A}\right]^n .
\end{eqnarray}
As the term $S^{(0)}_{\eff}$ does not depend on the gauge field, we must focus only on the term $S^{(1)}_{\eff}$. To generate the higher-derivative Chern-Simons term it is necessary to have two contributions of the gauge field. Thus, we take $n=2$ in the expression (\ref{seff1}) and use the following expansion for the fermionic propagator,
\begin{eqnarray} \frac{1}{\slashed{p}-m-\gamma_5\slashed{b}}&=&\frac{1}{\slashed{p}-m}+\frac{1}{\slashed{p}-m}\gamma_5\slashed{b}\frac{1}{\slashed{p}-m}\nonumber\\
&&+\frac{1}{\slashed{p}-m}\gamma_5\slashed{b}\frac{1}{\slashed{p}-m}\gamma_5\slashed{b}\frac{1}{\slashed{p}-m}+\cdots,
\end{eqnarray}
to single out contributions of first order in $b_\mu$. It is worth emphasizing that the higher-derivative Chern-Simons term (and the Myers-Pospelov term) is also induced when we consider contributions of third order in $b_\mu$, as it has been shown in~\cite{Mar2}. Following with the derivative expansion approach \cite{Ait,Fra,Vai,Bab,Baz}, we find
\begin{eqnarray}\label{Sb}
S_{b}^{(1,2)}&=&\frac{i}{2} \Tr [S(p)\gamma^\mu S(p-k)\gamma_5 \slashed{b}S(p-k)\gamma^\nu+S(p)\gamma_5 \slashed{b}S(p)\gamma^\mu S(p-k)\gamma^\nu ]A_\mu A_\nu,
\end{eqnarray}
where $S(p)=(\slashed{p}-m)^{-1}$. Finally, expanding the propagator up to third order in $k_\mu$,
\begin{eqnarray}
S(p-k)&=&S(p)+S(p)\slashed{k}S(p)+S(p)\slashed{k}S(p)\slashed{k}S(p)+S(p)\slashed{k}S(p)\slashed{k}S(p)\slashed{k}S(p)+\cdots,
\end{eqnarray}
we obtain
\begin{eqnarray}\label{S5}
S_{(5)}&=&\frac{i}{2}\int{\frac{d^4k}{(2 \pi)^4}\Pi^{\mu\nu}_{(5)}\, A_\mu(k) A_\nu(-k)},
\end{eqnarray}
with
\begin{eqnarray}\label{pi1}
\Pi^{\mu\nu}_{(5)} &=&\int\frac{d^4 p}{(2\pi)^4} \tr[S(p)\gamma^\mu S(p)\gamma_5 \slashed{b}S(p)\slashed{k} S(p)\slashed{k}S(p)\slashed{k} S(p)\gamma^\nu \nonumber\\
&&+ S(p)\gamma^\mu S(p)\slashed{k} S(p)\gamma_5 \slashed{b}S(p)\slashed{k} S(p)\slashed{k} S(p)\gamma^\nu \nonumber \\ 
&&+ S(p)\gamma^\mu S(p)\slashed{k} S(p)\slashed{k} S(p)\gamma_5 \slashed{b}S(p)\slashed{k} S(p)\gamma^\nu \nonumber\\ 
&&+ S(p)\gamma^\mu S(p)\slashed{k} S(p)\slashed{k} S(p)\slashed{k} S(p)\gamma_5 \slashed{b}S(p)\gamma^\nu  \nonumber \\
&&+ S(p)\gamma_5 \slashed{b}S(p)\gamma^\mu S(p)\slashed{k} S(p)\slashed{k} S(p)\slashed{k} S(p)\gamma^\nu],
\end{eqnarray}
where we have calculated the trace over the coordinate and momentum spaces and considered cubic terms of $k_\mu$. 

Before evaluating the integrals in~(\ref{pi1}), we first calculate the trace over the Dirac matrices, so that we have
\begin{eqnarray}\label{pi2} 
\Pi^{\mu\nu}_{(5)}&=& 4i \epsilon^{\mu \nu \sigma \rho} b_\sigma k_\rho k^2\int{ \frac{d^4 p}{(2\pi)^4}}\frac{1}{(p^2-m^2)^3} - 16i  \epsilon^{\mu \nu \sigma \rho} b_\sigma k_\rho k^\alpha k^\beta \int{ \frac{d^4 p}{(2\pi)^4}}\frac{p_\alpha p_\beta}{(p^2-m^2)^4}\nonumber \\
&&+ 24i m^2 \epsilon^{\mu \nu \sigma \rho}  b_\sigma k_\rho k^2 \int{ \frac{d^4 p}{(2\pi)^4}}\frac{1}{(p^2-m^2)^4} + 24i \epsilon^{\mu \nu \rho \alpha} k_\rho b^\beta k^2 \int{ \frac{d^4 p}{(2\pi)^4}}\frac{p_\alpha p_\beta}{(p^2-m^2)^4}\nonumber \\
&&+ 32i \epsilon^{\mu \nu \rho \alpha }k_\rho (b\cdot k)k^\beta \int{ \frac{d^4 p}{(2\pi)^4}}\frac{p_\alpha p_\beta}{(p^2-m^2)^4} - 128i \epsilon^{\mu \nu \sigma \rho} b_\sigma k_\rho k^\alpha k^\beta \int{ \frac{d^4 p}{(2\pi)^4}}\frac{p_\alpha p_\beta}{(p^2-m^2)^5}\nonumber\\
&&- 128i \epsilon^{\mu \nu \rho \alpha} k_\rho b^\beta k^\delta k^\gamma \int{ \frac{d^4 p}{(2\pi)^4}}\frac{p_\alpha p_\beta p_\delta p_\gamma}{(p^2-m^2)^5}. 
\end{eqnarray}
Now, by analyzing the above integrals we observe that, after a simple power counting, we do not need to use regularization schemes because all integrals are convergent. Therefore, we can solve (\ref{pi2}) directly by using well-known solutions in which the higher-derivative Chern-Simons term is given by
\begin{equation}\label{pi0}
\Pi^{\mu\nu}_{(5)}=-\frac{\epsilon^{\mu \nu \sigma \rho} b_\sigma k_\rho k^2}{12 m^2 \pi^2},
\end{equation}
or through the following Lagrangian as
\begin{equation}\label{lg0}
\mathcal{L}_{(5)}=\frac{e^2}{24 m^2 \pi^2} \epsilon^{\mu\nu\rho\sigma} b_\sigma A_\mu \partial_\rho \Box A_\nu,
\end{equation}
with $\mathcal{K}^{\mu \nu \rho \alpha \beta}_{(5)}= \frac{e^2}{24 m^2 \pi^2}\epsilon^{\mu\nu\sigma\rho} b_\sigma g^{\alpha \beta}$, where we have re-instaled the electron charge.

The higher-derivative Chern-Simons term (\ref{lg0}) is also induced through radiative corrections from another CPT-odd term of the SME, namely governed by the coefficient $g^{\mu\nu\rho}$, when $g^{\mu\nu\rho}$ is totally antisymmetric \cite{Mar}.

Numerical estimates for the coefficient $b_\mu$ of the higher-derivative Chern-Simons term can be obtained from bounds on the coefficient $\mathcal{K}^{\mu\nu\rho\alpha\beta}_{(5)}$ of the photon sector (see~\cite{Kos4}, table XIX). From observational data related to the cosmic microwave background (CMB) polarization, we can estimate $b\sim 10^{-24}$. Moreover, from systems related to the astrophysical birefringence, we estimate that the coefficient is $\sim 10^{-36}$. In these estimates we have considered $m$ as being the electron mass, $m\simeq0.5\times10^{-3}$GeV, and $e\simeq10^{-1}$ for the electron charge. In this way, we observe that we have found tiny values for the coefficient $b_\mu$ compatible with maximal sensitivities for the electron sector, table II of Ref.~\cite{Kos4}. 

\section{Radiative correction at finite temperature}
\label{finite}

In this section, we are interested in the finite temperature behavior of the higher-derivative term (\ref{pi0}). For this, we take the Eq.~(\ref{pi2}) and change it from Minkowski space to Euclidean space. We would like to emphasize here that the method of derivarive expansion used above, in general, is rather delicate at finite temperature, due to the fact that the limits $k_0\to0$ and $\vec k\to0$ do not commute, arising from the non-analyticity of thermal amplitudes \cite{Das}. In the expression~(\ref{pi2}) we have considered the limits $k_0\to0$ and $\vec k\to0$. Then, by performing the following procedure: $p_0 \rightarrow i p_0$ ($g^{\mu\nu}\rightarrow -\delta^{\mu\nu}$), $d^4p\rightarrow i d^4p_E$, and $p^2\rightarrow -p_0^2-\vec p^2 = -p^2_E$, we obtain
\begin{eqnarray}\label{pi3} 
\Pi^{\mu\nu}_{(5)}&=& - 4 \epsilon^{\mu \nu \sigma \rho} b^\sigma_E k^\rho_E k^2_E \int{\frac{d^4 p_E}{(2 \pi)^4}}\frac{1}{(p^2_E + m^2)^3} + 16 \epsilon^{\mu \nu \sigma \rho} b^\sigma_E k^\rho_E k^\alpha_E k^\beta_E \int{\frac{d^4 p_E}{(2 \pi)^4}}\frac{p^\alpha_E p^\beta_E}{(p^2_E + m^2)^4}\nonumber \\
 &&+ 24 m^2 \epsilon^{\mu \nu \sigma \rho}  b^\sigma_E k^\rho_E k^2_E \int{\frac{d^4 p_E}{(2 \pi)^4}}\frac{1}{(p^2_E + m^2)^4} - 24 \epsilon^{\mu \nu \rho \alpha} k^\rho_E b^\beta_E k^2_E \int{\frac{d^4 p_E}{(2 \pi)^4}}\frac{p^\alpha_E p^\beta_E}{(p^2_E + m^2)^4}\nonumber \\
 &&- 32 \epsilon^{\mu \nu \rho \alpha }k^\rho_E (b_E\cdot k_E)k^\beta_E \int{\frac{d^4 p_E}{(2 \pi)^4}}\frac{p^\alpha_E p^\beta_E}{(p^2_E + m^2)^4} - 128 \epsilon^{\mu \nu \sigma \rho} b^\sigma_E k^\rho_E k^\alpha_E k^\beta_E \int{\frac{d^4 p_E}{(2 \pi)^4}}\frac{p^\alpha_E p^\beta_E}{(p^2_E + m^2)^5} \nonumber \\
 &&+ 128 \epsilon^{\mu \nu \rho \alpha} k^\rho_E b^\beta_E k^\delta_E k^\gamma_E \int{\frac{d^4 p_E}{(2 \pi)^4}}\frac{p^\alpha_E p^\beta_E p^\delta_E p^\gamma_E}{(p^2_E + m^2)^5}. 
\end{eqnarray}

In order to effectively tackle the issue of the finite temperature behavior of the above expression, we separate the space and time components of the four-momentum $p^\sigma_E=(p_0,\vec{p})$ as $p^\sigma_E\rightarrow \hat{p}^\sigma +p_0\delta^{\sigma 0}$, so that $ \hat{p}^\sigma=(0,\vec{p})$. Also, due to the symmetry of the integral under spacial rotations, it is possible to use the substitutions
\begin{eqnarray}
\hat{p}^\alpha \hat{p}^\beta \rightarrow \frac{\hat{p}^2}{3}(\delta^{\alpha\beta}-\delta^{\alpha 0} \delta^{\beta 0})
\end{eqnarray}
and
\begin{eqnarray}
\hat{p}^\alpha \hat{p}^\beta \hat{p}^\delta \hat{p}^\gamma &\rightarrow& \frac{\hat{p}^4}{15}[(\delta^{\alpha\beta}-\delta^{\alpha 0}\delta^{\beta 0})(\delta^{\delta\gamma}-\delta^{\delta 0}\delta^{\gamma 0})+(\delta^{\alpha\delta}-\delta^{\alpha 0}\delta^{\delta 0})(\delta^{\beta\gamma}-\delta^{\beta 0}\delta^{\gamma 0})\nonumber\\ 
&&+(\delta^{\alpha\gamma}-\delta^{\alpha 0}\delta^{\gamma 0})(\delta^{\beta \delta}-\delta^{\beta 0}\delta^{\delta 0})].
\end{eqnarray}
After applying these changes into Eq.~(\ref{pi3}), we find seven different tensorial structures. The first two structures, $\epsilon^{\mu \nu \sigma \rho} k^{\sigma}_E k^{\rho}_E b^0_E k^0_E$ and $\epsilon^{\mu \nu \sigma \rho} k^{\sigma}_E k^{\rho}_E (b_E \cdot k_E)$, obviously vanish by antisymmetry of $\epsilon^{\mu \nu \sigma \rho}$. In addition, the third contribution, $\epsilon^{\mu \nu 0 \sigma} k^{0}_E k^{\sigma}_E (b_E \cdot k_E)$, goes to zero after the integration in the space components of $p_E^\alpha$. Therefore, we are left with four different structures to analyze, namely $\epsilon^{\mu \nu \sigma \rho} b^{\sigma}_E k^{\rho}_E k^2_E$, $\epsilon^{\mu \nu 0 \rho} b^{0}_E k^{\rho}_E k^2_E$, $\epsilon^{\mu \nu \sigma \rho} b^{\sigma}_E k^{\rho}_E (k^0_E)^2$ and $\epsilon^{\mu \nu 0 \rho} b^{0}_E k^{\rho}_E (k^0_E)^2$. Thus, we have $\Pi^{\mu\nu}_{(5)}=\Pi^{\mu\nu}_{(a)}+\Pi^{\mu\nu}_{(b)}+\Pi^{\mu\nu}_{(c)}+\Pi^{\mu\nu}_{(c)}$, with the following expressions for these four tensorial structures:
\begin{eqnarray}
\Pi^{\mu\nu}_{(a)}&=& -4 \epsilon^{\mu \nu \sigma \rho}  b^{\sigma}_E k^{\rho}_E k^2_E \int\frac{dp_0}{2 \pi}\left[ \frac{32}{15}I_1(p_0,m)+\frac{32 m^2}{3} I_2(p_0,m)-\frac{10}{3} I_4(p_0,m)\right. \nonumber\\
&&- 6 m^2 I_5(p_0,m)+ I_6(p_0,m)\bigg], \\
\Pi^{\mu\nu}_{(b)}&=& 8 \epsilon^{\mu \nu 0 \rho} b^{0}_E k^{\rho}_E k^2_E \int\frac{dp_0}{2 \pi} \left[ \frac{16}{15}I_1(p_0,m)-\frac{16}{3}p_0^2 I_2(p_0,m) - I_4(p_0,m)\right. \nonumber \\
&&\left.+ 3p_0^2 I_5(p_0,m)\right], \\
\Pi^{\mu\nu}_{(c)}&=& -16  \epsilon^{\mu \nu \sigma \rho} b^{\sigma}_E k^{\rho}_E  (k^0_E)^2\int\frac{dp_0}{2 \pi}\left[ -\frac{8}{15}I_1(p_0,m)+\frac{8}{3}(p_0^2-m^2)I_2(p_0,m) \right. \nonumber\\
&&\left.+8 p_0^2 m^2 I_3(p_0,m) + \frac{1}{3} I_4(p_0,m)-p_0^2I_5(p_0,m)\right], \\
\Pi^{\mu\nu}_{(d)}&=& -128\epsilon^{\mu \nu 0 \rho} b^{0}_E k^{\rho}_E (k^0_E)^2 \int\frac{dp_0}{2 \pi}\left[ \frac{1}{5}I_1(p_0,m)-2 p_0^2 I_2(p_0,m)+p_0^4 I_3(p_0,m)\right],
\end{eqnarray}
where the integrals $I_{1,2,3,4,5,6}(p_0,m)$ are given by
\begin{eqnarray}
I_{1,2,3}(p_0,m)&=& \int\frac{d^3p_E}{(2 \pi)^3}\frac{\alpha_{1,2,3}}{(\vec{p}^2+p_0^2+m^2)^5}, \\
I_{4,5}(p_0,m)&=& \int\frac{d^3p_E}{(2 \pi)^3}\frac{\alpha_{4,5}}{(\vec{p}^2+p_0^2+m^2)^4}, \\
I_6(p_0,m)&=& \int\frac{d^3p_E}{(2 \pi)^3}\frac{1}{(\vec{p}^2+p_0^2+m^2)^3},
\end{eqnarray}
with $\alpha_1=\vec{p}^4,\alpha_2=\alpha_4=\vec{p}^2$ and $\alpha_3=\alpha_5=1$.

Now, we calculate the integrals over the space components $\vec p$. This can be done without using regularization schemes, because all the integrals are convergent. Then, we find
\begin{eqnarray}
\Pi^{\mu\nu}_{(a)}&=& \frac{m^2}{ 8\pi}\epsilon^{\mu \nu \sigma \rho} b^{\sigma}_E k^{\rho}_E  k^2_E \left[\int\frac{dp_0}{2 \pi}(p_0^2+m^2)^{-\frac{5}{2}}-\frac{2}{3m^2}\int\frac{dp_0}{2 \pi}(p_0^2+m^2)^{-\frac{3}{2}}\right], \nonumber\\
&&+\frac{1}{12\pi}\epsilon^{\mu \nu \sigma \rho} b^{\sigma}_E k^{\rho}_E  k^2_E \int\frac{dp_0}{2 \pi}(p_0^2+m^2)^{-\frac{3}{2}} \\
\Pi^{\mu\nu}_{(b)} &=& -\frac{m^2}{8 \pi}\epsilon^{\mu \nu 0 \rho} b^{0}_E k^{\rho}_E k^2_E \left[\int\frac{dp_0}{2 \pi}(p_0^2+m^2)^{-\frac{5}{2}}-\frac{2}{3m^2}\int\frac{dp_0}{2 \pi}(p_0^2+m^2)^{-\frac{3}{2}}\right],\\
\Pi^{\mu\nu}_{(c)}&=& \frac{ m^2}{ \pi}\epsilon^{\mu \nu \sigma \rho} b^{\sigma}_E k^{\rho}_E (k^0_E)^2\left[\frac{5 m^2}{4} \int\frac{dp_0}{2 \pi}(p_0^2+m^2)^{-\frac{7}{2}}-\int\frac{dp_0}{2 \pi}(p_0^2+m^2)^{-\frac{5}{2}}\right], \\
\Pi^{\mu\nu}_{(d)}&=& -\frac{m^2}{\pi}\epsilon^{\mu \nu 0 \rho} b^{0}_E k^{\rho}_E (k^0_E)^2 \left[\frac{5 m^2}{4} \int\frac{dp_0}{2 \pi}(p_0^2+m^2)^{-\frac{7}{2}}-\int\frac{dp_0}{2 \pi}(p_0^2+m^2)^{-\frac{5}{2}}\right].
\end{eqnarray}
Note that $\Pi^{\mu\nu}_{(b)}$ and $\Pi^{\mu\nu}_{(d)}$ cancel the corresponding time components $b_E^0$ of $\Pi^{\mu\nu}_{(a)}$ and $\Pi^{\mu\nu}_{(c)}$, respectively. Therefore, by summing all expressions, we obtain
\begin{eqnarray}\label{piBC}
\Pi^{\mu\nu}_{(5)} &=& -\frac{1}{m^2}\epsilon^{\mu \nu \sigma \rho } b^{\sigma}_E k^{\rho}_E  k^2_E B(m) - \frac{1}{m^2}\epsilon^{\mu \nu i \rho} b^{i}_E k^{\rho}_E k^2_E C_1(m)\nonumber\\
&&- \frac{1}{m^2}\epsilon^{\mu \nu i \rho} b^{i}_E  k^{\rho}_E (k^0_E)^2 C_2(m),
\end{eqnarray}
where
\begin{subequations}
\begin{eqnarray}
B(m) &=& -\frac{m^2}{12 \pi} \int\frac{dp_0}{2 \pi}(p_0^2+m^2)^{-\frac{3}{2}},\\
C_1(m) &=& \frac{m^2}{12\pi}\int\frac{dp_0}{2 \pi}(p_0^2+m^2)^{-\frac{3}{2}} - \frac{m^4}{8 \pi}\int\frac{dp_0}{2 \pi}(p_0^2+m^2)^{-\frac{5}{2}}, \\
C_2(m) &=& \frac{m^4}{\pi}\int\frac{dp_0}{2 \pi}(p_0^2+m^2)^{-\frac{5}{2}} - \frac{5 m^6}{4 \pi} \int\frac{dp_0}{2 \pi}(p_0^2+m^2)^{-\frac{7}{2}},
\end{eqnarray}
\end{subequations}
with $i=1,2,3$.

The finite temperature behavior of the above terms can be obtained by using the Matsubara formalism. For this, we write $p_0=(n +1/2) \frac{2\pi}{\beta}$, which are the Matsubara frequencies, and change the integral over $p_0$ into a sum, $\frac{1}{2\pi}\int{dp_0}\rightarrow \frac{1}{\beta}\sum_n$, where $T=\beta^{-1}$ is the temperature of the system. Assuming that $\xi=\frac{\beta m}{2 \pi}$, we can write $B(m)\rightarrow B(\xi)$, $C_1(m)\rightarrow C_1(\xi)$, and $C_2(m)\rightarrow C_2(\xi)$, so that
\begin{subequations}\label{BCC1}
\begin{eqnarray}
\label{B}B(\xi) &=& -\frac{\xi^2}{24 \pi^2}\sum_n{[(n+1/2)^2+\xi^2]^{-3/2}},\\
\label{C1}C_1(\xi) &=& \frac{\xi^2}{24\pi^2}\sum_n{[(n+1/2)^2+\xi^2]^{-3/2}} - \frac{\xi^4}{16\pi^2} \sum_n{[(n+1/2)^2+\xi^2]^{-5/2}} \\
\label{C2}C_2(\xi) &=& \frac{\xi^4}{2 \pi^2} \sum_n{[(n+1/2)^2+\xi^2]^{-5/2}} - \frac{5 \xi^6}{8\pi^2}\sum_n{[(n+1/2)^2+\xi^2]^{-7/2}}.
\end{eqnarray}
\end{subequations}

As the above summations are all convergent, they can be easily numerically evaluated. Nevertheless, the asymptotic limits, zero and high temperatures, are not easy to obtain. One way to do this is to convert the sums into integrals, by using the expression~\cite{Ford}
\begin{eqnarray}\label{sum}
\sum\limits_{n}[(n+b)^2+a^2]^{-\lambda}&=&\frac{\sqrt{\pi}\Gamma(\lambda - 1/2)}{\Gamma(\lambda)(a^2)^{\lambda-1/2}}+ 4\sin(\pi \lambda) f_\lambda (a,b),
\end{eqnarray}
where 
\begin{eqnarray}\label{intsum}
f_\lambda(a,b)= \int\limits_{|a|}^{\infty} \frac{dz}{(z^2-a^2)^\lambda} Re \left( \frac{1}{e^{2\pi(z+ib)}-1} \right),
\end{eqnarray}
which is valid for $Re\,\lambda < 1$, asides from the poles at $\lambda = 1/2, -1/2, -3/2, \cdots.$ However, by analyzing the equations (\ref{B}), (\ref{C1}), and (\ref{C2}), we have the values $\lambda=3/2$, $\lambda=5/2$ and $\lambda=7/2$, i.e. they are all out of range of validity. Luckily, we can use a recurrence relation, given by
\begin{eqnarray}\label{rr}
f_{\lambda -1}(a,b)=\!-\frac{1}{2a^2}\frac{2\lambda-5}{\lambda-2}f_{\lambda-2}(a,b)-\frac{1}{4a^2}\frac{1}{(\lambda-3)(\lambda-2)}\frac{\partial^2}{\partial b^2}f_{\lambda-3}(a,b),
\end{eqnarray}
in order to decrease these values of $\lambda$. In this way, for $\lambda=3/2$ we must use the recurrence relation once, for $\lambda=5/2$  twice, and for $\lambda=7/2$ thrice. We must also write $\lambda=3/2\to D/2$, $\lambda=5/2\to D/2+1$, and $\lambda=7/2\to D/2+2$, to avoid the poles $\lambda=1/2,-1/2$ in the first term of the solution (\ref{sum}), so that the expressions (\ref{BCC1}) become
\begin{subequations}
\begin{eqnarray}
B(\xi) &=& -\frac{m^{3-D}}{24\pi^2}\xi^{D-1}\sum_n{[(n+1/2)^2+\xi^2]^{-\frac D2}},\\
C_1(\xi) &=& \frac{m^{3-D}}{24\pi^2}\xi^{D-1}\sum_n{[(n+1/2)^2+\xi^2]^{-\frac D2}} \nonumber \\
&&- \frac{m^{3-D}}{16\pi^2}\xi^{D+1}\sum_n{[(n+1/2)^2+\xi^2]^{-\frac D2-1}}, \\
C_2(\xi) &=&  \frac{m^{3-D}}{2 \pi^2}\xi^{D+1}\sum_n{[(n+1/2)^2+\xi^2]^{-\frac D2-1}} \nonumber\\
&&- \frac{5m^{3-D}}{8\pi^2}\xi^{D+3}\sum_n{[(n+1/2)^2+\xi^2]^{-\frac D2-2}}.
\end{eqnarray} 
\end{subequations}
Then, by using the recurrence relation (\ref{rr}) and considering the sum (\ref{sum}), after we take the limit $D\to3$, we obtain
\begin{eqnarray}\label{pift}
\Pi^{\mu\nu}_{(5)} &=&  \frac{1}{m^2}\epsilon^{\mu \nu \sigma \rho } b_{\sigma} k_{\rho} k^2 B(\xi) + \frac{1}{m^2} \epsilon^{\mu \nu i \rho } b_{i} k_{\rho}  k^2 C_1(\xi) \nonumber\\
&&+\frac{1}{m^2}\epsilon^{\mu \nu i \rho} b_{i} k_{\rho} k_0^2 C_2(\xi),
\end{eqnarray}
with
\begin{subequations}\label{BCC2}
\begin{eqnarray}
\label{G}B(\xi) &=& \frac{1}{6}\int\limits_{|\xi|}^{\infty}dz \sqrt{z^2-\xi^2}\tanh(\pi z)\sech^2(\pi z)-\frac{1}{12 \pi^2},\\
\label{H1}C_1(\xi) &=&\frac{\xi^2}{12}\int\limits_{|\xi|}^{\infty}dz\frac{ \tanh(\pi z)\sech^2(\pi z)}{ \sqrt{z^2-\xi^2}}, \\
\label{H2}C_2(\xi) &=& \frac{\pi^2 \xi^2}{6}\int\limits_{|\xi|}^{\infty}dz\sqrt{z^2-\xi^2}\sech^5(\pi z)[\sinh(3\pi z)-11\sinh(\pi z)],
\end{eqnarray}
\end{subequations}
where we have returned to the Minkowski space.

By analyzing the above equations we observe that when $T\to0$ ($\xi\to\infty$) all the integrals vanish, so that $\Pi^{\mu\nu}_{(5)}$ goes to $-\frac{1}{12 \pi^2 m^2}\epsilon^{\mu \nu \sigma \rho } b_{\sigma} k_{\rho} k^2$. This is in fact the result of the temperature zero, obtained previously in~(\ref{pi0}). On the other hand, when $T\to\infty$ ($\xi\to0$) the expression (\ref{G}) goes to zero, as well as the expressions (\ref{H1}) and (\ref{H2}). This happens mainly because all the summations in (\ref{BCC1}) are strongly suppressed by the temperature, thus suggesting that in general the operators of mass dimension $d\ge5$ vanish in the limit of high temperature.

\begin{figure}[h]
\includegraphics[scale=.8]{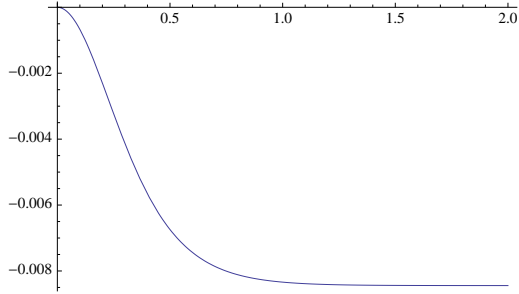}
\caption{Plot of the function $B(\xi)$} \label{fig1}
\end{figure}

\begin{figure}[h]
\includegraphics[scale=.7]{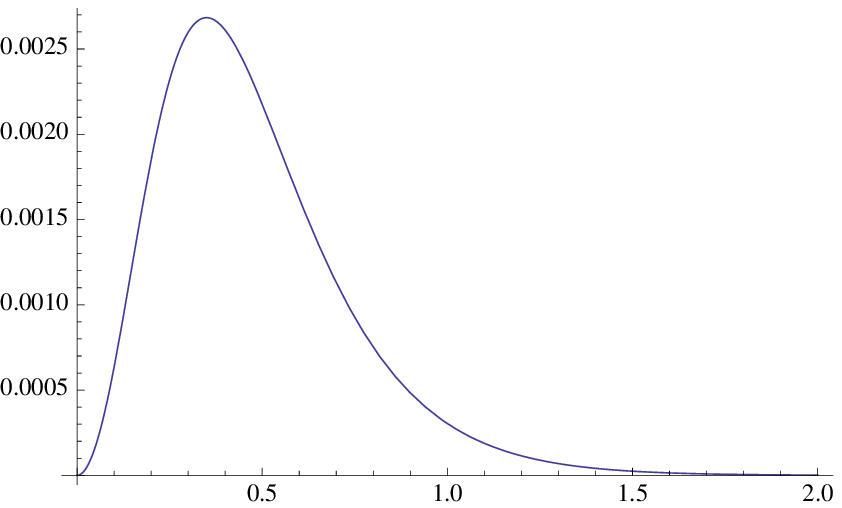}
\caption{Plot of the function $C_1(\xi)$} \label{fig2}
\end{figure}

\begin{figure}[h]
\includegraphics[scale=.7]{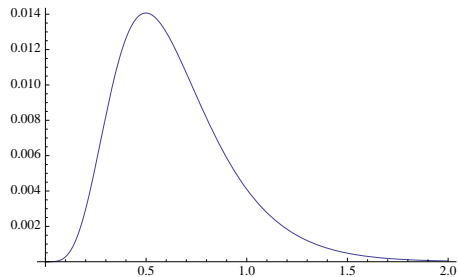}
\caption{Plot of the function $C_2(\xi)$} \label{fig3}
\end{figure}

The plot of the functions $B(\xi)$, $C_1(\xi)$, and $C_2(\xi)$, which can be numerically calculated from either expressions (\ref{BCC1}) or (\ref{BCC2}), are presented in Figs.~\ref{fig1},~\ref{fig2}, and~\ref{fig3}. It is interesting to note here that these functions are the same as those obtained in the context of the coefficient $g^{\mu\nu\rho}$ \cite{Lei}. This is because under a certain fermion field redefinition \cite{Kos2,McD} the totally antisymmetric component of $g^{\mu\nu\rho}$ is completely absorbed into coefficient $b^\mu$.


\section{Large gauge invariance}
\label{large}

Let us now briefly discuss the question of large gauge invariance, at finite temperature, in the context of Lorentz- and CPT-violating QED. This issue has been extensively studied in three-dimensional Chern-Simons theories \cite{Dun,Des1,Des2,Fos1,Fos2,Fos3,Bra1,Bra2}. In our case, we want to examine the invariance of the higher-derivative Chern-Simons term (\ref{hd5}) under large gauge transformation, as well as the four-dimensional Chern-Simons term, namely,
\begin{eqnarray}\label{hd3}
S_{(3)}=\int{d^4x}\, \mathcal{K}^{\mu \nu \rho}_{(3)} A_{\mu}\partial_{\rho} A_{\nu},
\end{eqnarray}
with $\mathcal{K}^{\mu \nu \rho}_{(3)}\propto \epsilon^{\mu \nu \rho \sigma} b_\sigma$, where both are dynamically generated by the effective action (\ref{seff}). It would be interesting to consider the other terms with operators of mass dimension $d=7,9,\cdots$, however, this is a challenging task. In fact, in the literature we find only models with Lorentz-violating operators of mass dimension $d=5$, for CPT-odd terms. In particular, we can argue that, as ${\cal K}_{(d)}\sim M_{Pl}^{4-d}$, these operators with $d\geq 7$ are highly suppressed by the mass Planck $M_{Pl}$ and, therefore, are even more negligible.

To perform this investigation, we continue to use the method of derivative expansion, however, in a particular gauge field background, where we have a vanishing electric field and a time-independent magnetic field, as follows
\begin{equation}\label{bg}
A_0=A_0(t), \hspace{0.5cm} A_i=A_i(x,y,z). 
\end{equation}
It is worth mentioning that this specific choice of gauge background is consistent with the static limit, as shown in~\cite{Bra1}, although it is not a static background. The point here is that in this static limit we can avoid the non-analyticity of thermal amplitudes, which can appear when the effective action is expanded in powers of derivatives. Furthermore, in this background (\ref{bg}) we believe that the effective action can be calculated exactly. 

Finally, by considering the appropriate gauge transformation \cite{Fos1,Fos2}, 
\begin{equation}
A_\mu \to A_\mu + \partial_\mu \Omega, \hspace{0.5cm} \Omega(t)=\left(-\int_0^t+\frac t\beta\int_0^\beta\right)dt' A_0(t'),
\end{equation}
we obtain
\begin{equation}
A_0 \to \frac{a_0}\beta = \frac1\beta \int_0^\beta dt\, A_0(t), \hspace{0.5cm} A_i \to A_i(x,y,z),
\end{equation}
so that under large gauge transformation,
\begin{equation}\label{lgt}
a_0 \to a_0 + 2\pi n,
\end{equation}
where $n$ is an integer. Then, by taking into account these constraints, we can rewrite the effective action (\ref{seff}) in the Euclidean space as
\begin{equation}\label{seffe}
S_{\eff}=-\sum_n\Tr\ln(\vec{\slashed{p}}+\tilde\omega_n\gamma^0+m-\vec{\slashed{A}}-\gamma_5 \slashed{b}_E),
\end{equation} 
in which we have introduced the definitions: $\vec{\slashed{p}}=p^i\gamma^i$, $\slashed{b}_E=b^0\gamma^0+b^i\gamma^i$,  and
\begin{equation}
\tilde\omega_n=\omega_n-\frac{a_0}\beta=(n +1/2) \frac{2\pi}{\beta}-\frac{a_0}\beta.
\end{equation}
Now, following the procedure performed in Ref.~\cite{Fos3}, let us first derivate the effective action (\ref{seffe}) with respect to $a_0$, and later consider only the linear term in $A^i$, so that we can write
\begin{equation}
\frac{\partial S_{\eff}^{(1)}}{\partial a_0} = \sum_n \Tr\, \frac{1}{\vec{\slashed{p}}+\tilde\omega_n\gamma^0+m-\gamma_5 \slashed{b}_E}\vec{\slashed{A}}  \frac{1}{\vec{\slashed{p}}+\tilde\omega_n\gamma^0+m-\gamma_5 \slashed{b}_E} \gamma^0.
\end{equation}
As we are interested only in contributions of first order in $b_\mu$, we single out the expression
\begin{equation}\label{Sb1}
\frac{\partial S_b^{(1)}}{\partial a_0} = \sum_n \Tr\, [S_E(\vec p)\gamma^i S_E(\vec p-i\vec\partial)\gamma_5 \slashed{b}_E S_E(\vec p-i\vec\partial)\gamma^0+S_E(\vec p)\gamma_5 \slashed{b}_E S_E(\vec p)\gamma^i S_E(\vec p-i\vec\partial)\gamma^0 ]A^i,
\end{equation}
where $S_E(\vec p)=(\vec{\slashed{p}}+\tilde\omega_n\gamma^0+m)^{-1}$. Note that the above equation is similar to Eq.~(\ref{Sb}), and therefore the calculations of Eq.~(\ref{Sb1}) are similar to those that have been performed previously.

In order to generate the Lorentz-violating Chern-Simons term (\ref{hd3}), we expand the propagator $S_E(\vec p-i\vec\partial)$ up to first order in $\partial^k$, so that we obtain
\begin{equation}
\frac{\partial S_{(3)}}{\partial a_0} = i \int d^3x\, \Pi_{(3)}^{i0} A^i
\end{equation}
with
\begin{eqnarray}\label{pi1}
\Pi_{(3)}^{i0} &=& \sum_n \int\frac{d^3 p_E}{(2\pi)^3} \tr[S_E(\vec p)\gamma_5 \slashed{b}_E S_E(\vec p)\gamma^i S_E(\vec p)\vec{\slashed{\partial}} S_E(\vec p)\gamma^0 \nonumber\\
&&+ S_E(\vec p)\gamma^i S_E(\vec p)\vec{\slashed{\partial}} S_E(\vec p)\gamma_5 \slashed{b}_E S_E(\vec p)\gamma^0 \nonumber \\ 
&&+ S_E(\vec p)\gamma^i S_E(\vec p)\gamma_5 \slashed{b}_E S_E(\vec p)\vec{\slashed{\partial}} S_E(\vec p)\gamma^0].
\end{eqnarray}
Hereafter, we calculate the trace over the Dirac matrices and thus find the expression
\begin{eqnarray}
\Pi_{(3)}^{i0} &=& -12i \epsilon^{i0jk} b^j \partial^k \sum_n\int\frac{d^3 p_E}{(2 \pi)^3}\frac{1}{(\vec p^2 + \tilde\omega_n^2 + m^2)^2} \nonumber\\
&&-16i \epsilon^{0jkl} b^j \partial^k \sum_n\int\frac{d^3 p_E}{(2 \pi)^3}\frac{p^ip^l}{(\vec p^2 + \tilde\omega_n^2 + m^2)^3} \nonumber\\
&&+16i \epsilon^{ijk0} b^j \partial^k \sum_n\int\frac{d^3 p_E}{(2 \pi)^3}\frac{\tilde\omega_n^2}{(\vec p^2 + \tilde\omega_n^2 + m^2)^3} \nonumber\\
&&+16i \epsilon^{i0jl} b^j \partial^k \sum_n\int\frac{d^3 p_E}{(2 \pi)^3}\frac{p^k p^l}{(\vec p^2 + \tilde\omega_n^2 + m^2)^3}, 
\end{eqnarray}
where we have moved $\gamma_5$ to the very end of every expression, before calculating the trace. The next step is to calculate the space integral, by promoting the three- to D-dimension and using the substitution $p^ip^l\to \vec p^2\delta^{il}/D$, which leads to the following result:
\begin{eqnarray}\label{Pi3}
\Pi_{(3)}^{i0} &=&  -i\epsilon^{i0jk} b^j \partial^k 2^{1-D}m\,\pi^{-D/2}(\mu^2)^{\frac32-\frac D2} \Gamma\left(2-\frac D2\right)\nonumber\\
&&\times \sum_n\left[\frac{(D-3)}{(\tilde\omega_n^2+m^2)^{2-\frac D2}}-\frac{(D-4)m^2}{(\tilde\omega_n^2+m^2)^{3-\frac D2}}\right].
\end{eqnarray}
To perform the above summations, we cannot readily take the limit $D\to3$ in the expression because the first sum exhibits singularity. Therefore, we must use the Eq.~(\ref{sum}) in order to isolate the corresponding divergent term, which becomes finite due to the factor $(D-3)$ in the numerator. For the second sum, when $D\to3$, we have $\lambda \to 3/2$, i.e. we also want to use here the recurrence relation (\ref{rr}). Following these procedures, we arrive at
\begin{equation}
\Pi_{(3)}^{i0} = -i\epsilon^{i0jk} b^j \partial^k\,F'(\xi,a_0),
\end{equation}
where
\begin{equation}
F'(\xi,a_0) = \int_{|\xi|}^\infty dz \sqrt{z^2-\xi^2}\,\sinh(2\pi z)\frac{2\cos(a_0)\cosh(2\pi z)-\cos(2a_0)+3}{[\cos(a_0)+\cosh(2\pi z)]^3},
\end{equation}
or, in other words,
\begin{equation}\label{F}
F'(\xi,a_0) = \frac{\partial\;}{\partial a_0}F(\xi,a_0) = \frac{\partial\;}{\partial a_0} \int_{|\xi|}^\infty dz \sqrt{z^2-\xi^2}\,\frac{2\sin(a_0)\sinh(2\pi z)}{[\cos(a_0)+\cosh(2\pi z)]^2}.
\end{equation}
Therefore, after returning to the Minkowski space, we obtain the Lorentz-violating Chern-Simons action,
\begin{equation}\label{S3lgt}
S_{(3)} = \int d^3x\, \epsilon^{i0jk} b_j \partial_k A_i\, F(\xi,a_0),
\end{equation}
with
\begin{equation}
F(\xi,a_0) = \int_{|\xi|}^\infty dz \sqrt{z^2-\xi^2}\,\frac{2\sin(a_0)\sinh(2\pi z)}{[\cos(a_0)+\cosh(2\pi z)]^2},
\end{equation}
which, as expected, is invariant under large gauge transformation (\ref{lgt}). Note that a perturbative expansion in terms of $a_0$ yields the usual perturbative result,
\begin{equation}
F(\xi,a_0) = a_0 \int_{|\xi|}^\infty dz \sqrt{z^2-\xi^2}\,\sech^2(\pi z)\tanh(\pi z) + {\cal O}(a_0^2),
\end{equation}
that it was obtained previously in~\cite{Cer}, and in~\cite{Lei} from the derivative term $\bar\psi {i\over2}g^{\kappa\lambda\mu}\sigma_{\kappa\lambda}(\partial_\mu+iA_\mu)\psi$, governed by the coefficient $g^{\kappa\lambda\mu}$.

Let us finally consider the generation of the higher-derivative Chern-Simons term (\ref{hd5}), by expanding the propagator $S_E(\vec p-i\vec\partial)$, in Eq.~(\ref{Sb1}), up to three order in $\partial^k$. As we have mentioned, the corresponding action is similar to Eq.~(\ref{S5}), however, given by the expression
\begin{equation}
\frac{\partial S_{(5)}}{\partial a_0} = i \int d^3x\, \Pi_{(5)}^{i0} A^i,
\end{equation}
where
\begin{eqnarray} 
\Pi^{i0}_{(5)}&=& 4i \epsilon^{i0jk} b^j \partial^k \vec\nabla^2 \sum_n \int{\frac{d^3 p_E}{(2 \pi)^3}}\frac{1}{(\vec p^2 + \tilde\omega_n^2 + m^2)^3} \nonumber\\
&&- 16i \epsilon^{i0jk} b^j \partial^k \partial^l \partial^m \sum_n \int{\frac{d^3 p_E}{(2 \pi)^3}}\frac{p^l p^m}{(\vec p^2 + \tilde\omega_n^2 + m^2)^4}\nonumber \\
 &&- 24i m^2 \epsilon^{i0jk}  b^j \partial^k \vec\nabla^2 \sum_n \int{\frac{d^3 p_E}{(2 \pi)^3}}\frac{1}{(\vec p^2 + \tilde\omega_n^2 + m^2)^4} \nonumber\\
 &&- 24i \epsilon^{i0jk} b^l \partial^k \vec\nabla^2 \sum_n \int{\frac{d^3 p_E}{(2 \pi)^3}}\frac{p^k p^l}{(\vec p^2 + \tilde\omega_n^2 + m^2)^4}\nonumber \\
&&+ 128i \epsilon^{i0jk} b^j \partial^k \partial^l \partial^m \sum_n \int{\frac{d^3 p_E}{(2 \pi)^3}}\frac{p^l p^m}{(\vec p^2 + \tilde\omega_n^2 + m^2)^5} \nonumber \\
 &&- 128i \epsilon^{i0kl} b^j \partial^k  \partial^m \partial^n \sum_n \int{\frac{d^4 p_E}{(2 \pi)^4}}\frac{p^j p^l p^m p^n}{(\vec p^2 + \tilde\omega_n^2 + m^2)^5}.
\end{eqnarray}
In the above expression we must also use the substitution $p^j p^l p^m p^n \to \vec p^4(\delta^{jl}\delta^{mn}+\delta^{jm}\delta^{ln}+\delta^{jn}\delta^{ml})/[D(D+2)]$, so that after we evaluate the space integrals and the sums, we obtain 
\begin{equation}
\Pi^{i0}_{(5)} = -\frac{i}{m^2}\epsilon^{i0jk} b^j \partial^k \vec\nabla^2 G'(\xi,a_0),
\end{equation}
with
\begin{equation}
G'(\xi,a_0) = \frac1{12}\int_{|\xi|}^\infty dz \frac{2z^2-\xi^2}{\sqrt{z^2-\xi^2}}\,\sinh(2\pi z)\frac{2\cos(a_0)\cosh(2\pi z)-\cos(2a_0)+3}{[\cos(a_0)+\cosh(2\pi z)]^3} -\frac{1}{12 \pi^2},
\end{equation}
which can be written as
\begin{equation}
G'(\xi,a_0) = \frac{\partial\;}{\partial a_0}G(\xi,a_0) = \frac{\partial\;}{\partial a_0} \left(\frac1{12}\int_{|\xi|}^\infty dz \frac{2z^2-\xi^2}{\sqrt{z^2-\xi^2}}\,\frac{2\sin(a_0)\sinh(2\pi z)}{[\cos(a_0)+\cosh(2\pi z)]^2}-\frac{a_0}{12 \pi^2}\right).
\end{equation}
Then, the higher-derivative Chern-Simons action takes the form
\begin{equation}\label{S5lgt}
S_{(5)} = \int d^3x\, \epsilon^{i0jk} b_j \partial_k \vec\nabla^2A_i\, G(\xi,a_0),
\end{equation}
where
\begin{equation}
G(\xi,a_0) = \frac1{12}\int_{|\xi|}^\infty dz \frac{2z^2-\xi^2}{\sqrt{z^2-\xi^2}}\,\frac{2\sin(a_0)\sinh(2\pi z)}{[\cos(a_0)+\cosh(2\pi z)]^2}-\frac{a_0}{12 \pi^2},
\end{equation}
which is also invariant under large gauge transformation (\ref{lgt}). Observe that by expanding the above expression, up to first order in $a_0$, Eqs.~(\ref{G}) and (\ref{H1}) are readily recovered. This complete our analysis.

\section{Summary}
\label{summary}

In this work, we study the radiative generation of the higher-derivative Lorentz-violating Chern-Simons term, at zero temperature and at finite temperature. At zero temperature, we use the method of derivative expansion, as a preparation for the finite temperature analysis. We also present numerical estimations for the coefficient $b^\mu$, which are compatible with maximal sensitivities for the electron sector. At finite temperature, by using the Matsubara formalism, we observe that the higher-derivative Chern-Simons term vanishes when the temperature goes to infinity. This behavior has already been observed in the context of the coefficient $g^{\mu\nu\rho}$ \cite{Lei}, suggesting, therefore, that in general the operators of mass dimension $d\ge5$ vanish in the limit of $T\to\infty$. In the analysis of large gauge invariance, we compute the exact effective actions, Eqs.~(\ref{S3lgt}) and (\ref{S5lgt}), at finite temperature and in a specific gauge field background. We observe that, in fact, the conventional Chern-Simons (\ref{S3lgt}) and the higher-derivative Chern-Simons (\ref{S5lgt}) actions are invariant under large gauge transformation. 

\vspace{.5cm}
{\bf Acknowledgements.} This work was supported by Conselho Nacional de Desenvolvimento Cient\'{\i}fico e Tecnol\'{o}gico (CNPq).

\end{document}